\documentclass[prl,twocolumn,showpacs,superscriptaddress]{revtex4}% Physical Review B

\usepackage{graphicx}% Include figure files
\usepackage{bm}% bold math

\newcommand{\beq}{\begin{equation}}
\newcommand{\eeq}{\end{equation}}
\newcommand{\beqa}{\begin{eqnarray}}
\newcommand{\eeqa}{\end{eqnarray}}

\newcommand{\w}{\omega}

\newcommand{\ket}[1]{\left| #1 \right\rangle}
\newcommand{\bra}[1]{\left\langle #1 \right|}

\begin{document}

\title{Selective Spin Coupling through a Single Exciton}
\author{A.~Nazir}
\email{ahsan.nazir@materials.oxford.ac.uk}
\affiliation{Department of Materials, Oxford University, Oxford OX1 3PH, United Kingdom}
\author{B.~W.~Lovett}
\email{brendon.lovett@materials.oxford.ac.uk}
\affiliation{Department of Materials, Oxford University, Oxford OX1 3PH, United Kingdom}
\author{S.~D.~Barrett}
\affiliation{Hewlett-Packard Laboratories, Filton Road,
Stoke Gifford, Bristol BS34 8QZ, United Kingdom}
\author{T.~P.~Spiller}
\affiliation{Hewlett-Packard Laboratories, Filton Road,
Stoke Gifford, Bristol BS34 8QZ, United Kingdom}
\author{G.~A.~D.~Briggs}
\affiliation{Department of Materials, Oxford University, Oxford OX1 3PH, United Kingdom}
%\date{\today}
% It is always \today, today, but any date may be explicitly specified
\begin{abstract}
We present a novel scheme for performing a conditional phase gate between two spin qubits in adjacent semiconductor
quantum dots through delocalized single exciton states, formed through the inter-dot F\"orster interaction. We consider
two resonant quantum dots, each containing a single excess conduction band electron whose spin embodies the qubit. We
demonstrate that both the two-qubit gate, and arbitrary single-qubit rotations, may be realized to a high fidelity with
current semiconductor and laser technology.
\end{abstract}


\maketitle

%\section{Introduction}
Quantum information processors could provide us with revolutionary algorithms for a wide range of
appications~\cite{nielsen00}.
Until recently, schemes for the implementation of such devices within semiconductor quantum dots
(QDs) usually fell into distinct categories based upon their proposed qubit. For example, quantum gates based on exciton
qubits~\cite{troiani00,biolatti02,lovett03,quiroga99} and spin qubits~\cite{loss98} have
been put forward. However, a new field of `hybrid' schemes is now emerging with the aim of marrying the advantageous
aspects of these individual candidate systems~\cite{imamoglu99,pazy03,feng04}.

In this Letter, we analyse the possibility of all-optical selective coupling of electron spins in adjacent QDs via
intermediate excitonic states. Motivated by the work of Refs.~{[\onlinecite{pazy03,feng04}]}, where the static dipole-dipole
interactions between two excitonic states were exploited, our scheme also benefits from the fast (picosecond) time scales
of excitonic interactions, along with the relative stability of spin qubits to decoherence~\cite{hanson03,golovach03}.
In contrast to the previous work, we consider the inter-dot resonant energy transfer
(F\"orster) interaction~\cite{lovett03} and find that a simple two qubit gate requires the excitation of
\emph{single exciton states only}.
This requires only one laser pulse, and so we
believe that it may be more readily implemented with current semiconductor and
laser technology, and that it represents a significant step towards a working spin-based optical quantum logic gate.

We consider two resonant QDs, each of which are $n$-doped so that they
contain a single excess conduction band electron~\cite{cortez02}.
The qubit basis
$|0\rangle$ and $|1\rangle$ is defined by the electron spin states $m_{z}=-1/2$ and $1/2$ respectively.
Ideally, we would like to implement a controlled phase (CPHASE) gate given by:
$  \{|00\rangle\rightarrow|00\rangle,
  |01\rangle\rightarrow|01\rangle,
  |10\rangle\rightarrow|10\rangle$
and  $|11\rangle\rightarrow-|11\rangle\}$.
To perform this operation, we consider a single laser radiating both QDs with $\sigma^{+}$ polarized light, resonant with
the {\it {s}}-shell heavy-hole exciton creation energies (the formulation for light-holes is equivalent).
These excitons are necessarily in the $z$-angular momentum states
given by $|3/2^{hh},-1/2^{e}\rangle$. As noted in
Refs.~{[\onlinecite{pazy03,feng04}]}, such a polarized laser will  create an exciton on a QD only if its excess electron
is in the spin state $|m_{z}=1/2\rangle$, due to the Pauli blocking effect for the two conduction band electrons. We
denote such a combined exciton-spin (trion) state by $|{X}\rangle$.

The Hamiltonian for our two resonantly coupled QDs interacting with a single
classical laser field may be written as
%{\setlength\arraycolsep{0.0em}
\begin{eqnarray}\label{spinexH}
H(t)&{}={}&\omega_a(|{X}\rangle\langle{X}|\otimes\hat{I}+\hat{I}\otimes|{X}\rangle\langle{X}|)\nonumber\\
&&\:{+}V_{XX}|{X}{ X}\rangle\langle{X}{X}|+V_{F}(|1{ X}\rangle\langle{X}1|+{\rm H.c.})\nonumber\\
&&\:{+}\Omega\cos{\omega_{l}t}(|1\rangle\langle{X}|\otimes\hat{I}+\hat{I}\otimes|1\rangle\langle{X}|+{\rm H.c.}),\nonumber\\
\end{eqnarray}
where H.c. denotes the Hermitian conjugate.
$\omega_{a}$ is the exciton creation energy for each dot, $V_{{F}}$ is the inter-dot F\"orster coupling strength,
$V_{{XX}}$ is the biexcitonic energy shift due to the exciton-exciton dipole interaction~\cite{biolatti02},
$\Omega$ is the time-dependent
coupling between laser and dot (taken to be the same for both QDs), and $\omega_{l}$ is the laser frequency. We have
assumed that the energy difference between states $|0\rangle$ and $|1\rangle$ is negligible on the exciton energy scales.
Single-particle tunneling between the dots is also neglected. The frequency dependence of the F\"orster
interaction~\cite{citrin95} can also be ignored for this two dot case, where the separation of the qubits is much smaller
than the wavelength of any emitted light; these
effects may, however, become important for a chain of dots in a scaled up device.

Our QDs are assumed to be smaller than the bulk exciton radius and hence within the strong confinement regime. Therefore,
the mixing of single-particle electron and hole states due their Coulomb interactions may be neglected, and any energy shift
can be incorporated into $\omega_{a}$~\cite{schmittrink87}. However, the Coulomb energies are still important as they
ensure that the resonance condition for single-particle tunneling is not the same as that for resonant exciton transfer.

As a consequence of the Pauli exclusion principle, there are no matrix elements that can cause transitions between
$|0\rangle$ and $|{X}\rangle$ on either dot. Additionally, to first order, the (non-magnetic) F\"orster process
couples only the states $|1{X}\rangle$ and $|{X}1\rangle$ and exchanges no spin information; hence single
delocalized excitons may exist only on pairs of dots {\it in the same spin state}, and this is the essence of our scheme.

We can see that the Hamiltonian of Eq.~\ref{spinexH} may be decoupled into four separate subspaces with no interactions
between them: $\{|00\rangle\},\{|01\rangle,|0{X}\rangle\},\{|10\rangle,|{X}0\rangle\},\{|11\rangle,|1{X}\rangle,
|{X}1\rangle,|{XX}\rangle\}$. We are primarily interested in the dynamics of the
$\{|11\rangle,|1{X}\rangle,|{X}1\rangle,|{XX}\rangle\}$ subspace where the coupling between the states
$|11\rangle,|1{X}\rangle,$ and $|{X}1\rangle$ will allow us to generate a phase shift on the input state
$|11\rangle$.
%
%In this subspace the Hamiltonian is:
%\begin{equation}\label{subspaceH}
%H_{sub}(t)=\left(
%\begin{array}{cccc}
 % \langle11| & \langle1\rm{X}| & \langle\rm{X}1| & \langle\rm{XX}| \\
 % 0 & \Omega\cos{\omega_{l}t} & \Omega\cos{\omega_{l}t} & 0\\
  %\Omega\cos{\omega_{l}t} & \omega_{a} & V_{\rm{F}} & \Omega\cos{\omega_{l}t}\\
  %\Omega\cos{\omega_{l}t} & V_{\rm{F}} & \omega_{a} & \Omega\cos{\omega_{l}t}\\
  %0 & \Omega\cos{\omega_{l}t} & \Omega\cos{\omega_{l}t} & 2\omega_{a}+V_{\rm {XX}}\\
 %\end{array}\right).
%\end{equation}
As we are considering the case of two resonant QDs it is sensible to first rewrite the Hamiltonian for this subspace
in a basis of its eigenstates when $\Omega=0$. These are
$\ket{11}$,
$|\psi_{+}\rangle=2^{-1/2}(|1{X}\rangle+|{X}1\rangle)$,
$|\psi_{-}\rangle=2^{-1/2}(|1{X}\rangle-|{X}1\rangle)$ and
$\ket{XX}$.
We obtain
\begin{eqnarray}\label{diagonalHlaser}
H_{\rm sub}(t)&=&(\w_a+V_F) \ket{\psi_+}\bra{\psi_+}+(\w_a-V_F)\ket{\psi_-}\bra{\psi_-}\nonumber\\
&&+(2\w_a+V_{XX})\ket{XX}\bra{XX}\nonumber\\
&&+\Omega'\cos{\omega_{l}t}(\ket{11}\bra{\psi_+} + \ket{\psi_+}\bra{XX}+{\rm H.c.}),\nonumber\\
&&
\end{eqnarray}
%
%\left(
%\begin{array}{cccc}
 % 0 & \Omega'\cos{\omega_{l}t} & 0 & 0 \\
 % \Omega'\cos{\omega_{l}t} & \omega_{a}+V_{\rm{F}} & 0 & \Omega'\cos{\omega_{l}t} \\
 % 0 & 0 & \omega_{a}-V_{\rm{F}} & 0 \\
 % 0 & \Omega'\cos{\omega_{l}t} & 0 & 2\omega_{a}+V_{\rm {XX}} \\
 %\end{array}
 %\right),
%\end{equation}
where the only dipole allowed transitions are between the states $|11\rangle$ and $|\psi_{+}\rangle$, and
between $|\psi_{+}\rangle$ and $|{XX}\rangle$, and $\Omega'=\sqrt{2}\Omega$.

\begin{figure}[t]
\centering
\includegraphics[width=3.3in,height=1.8in]{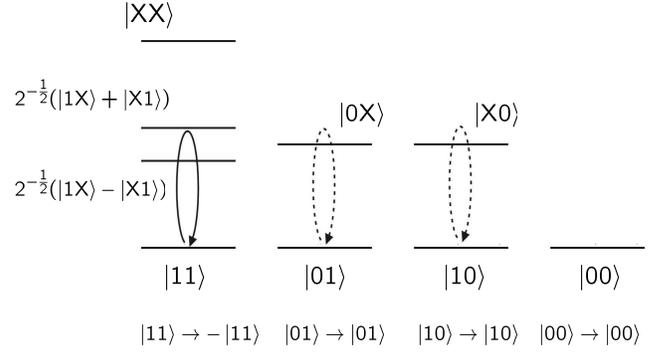}
\caption{Schematic of the energy level structure of two resonantly coupled QDs, showing its dependence on the spin state
of the excess electron on each dot. When the spins are both in state $\ket{1}$ an energy shift of the single
exciton states occurs due to their F\"orster coupling. Hence a phase shift may be accumulated using a laser resonant with
the dipole allowed transition. This does not occur for any of the other states shown due to the Pauli blocking mechanism.}
\label{scheme}
\end{figure}

We shall now transform this Hamiltonian into a frame rotating with the laser frequency $\omega_{l}$ with respect to both
dipole allowed transitions.
Within the rotating wave approximation, Eq.~\ref{diagonalHlaser} becomes
\begin{eqnarray}\label{Hrwaresonant}
H'_{\rm sub}(t)&=&-2V_F\ket{\psi_-}\bra{\psi_-}+(V_{XX}-2V_F)\ket{XX}\bra{XX}\nonumber\\
&&+\frac{\Omega'}{2}(\ket{11}\bra{\psi_+} + \ket{\psi_+}\bra{XX}+{\rm H.c.}),
\end{eqnarray}
%H'_{sub}=\left(
%\begin{array}{cccc}
 % 0 & \Omega'/2 & 0 & 0 \\
  %\Omega'/2 & 0 & 0 & \Omega'/2 \\
  %0 & 0 & -2V_{\rm{F}} & 0 \\
  %0 & \Omega'/2 & 0 & V_{\rm {XX}}-2V_{\rm{F}} \\
 %\end{array}
% 5\right),
%\end{equation}
where we set our laser frequency to be resonant with the dipole allowed
$|11\rangle$ to $|\psi_{+}\rangle$ transition; i.e., we set $\omega_{l}=(\omega_{a}+V_{{F}})$,
as this allows us to generate the desired CPHASE gate (see Fig.~\ref{scheme}).
Under the condition
\begin{equation}\label{condition1}
|\Omega'|/2\ll|V_{{XX}}-2V_{{F}}|,
\end{equation}
the double excitation of $\ket{11}$ to $\ket{XX}$ is suppressed, and
we may use second order degenerate perturbation theory to decouple the two subspaces $\{|11\rangle,|\psi_{+}\rangle\}$ and
$\{|\psi_{-}\rangle,|{XX}\rangle\}$.
By doing this, we may write down an effective Hamiltonian in the degenerate subspace
$\{|11\rangle,|\psi_{+}\rangle\}$ as
\begin{eqnarray}\label{Hsubeff}
H_{\rm eff}(t)&=&\frac{-\Omega'^2}{4(V_{XX}-2V_F)}\ket{\psi_+}\bra{\psi_+}\nonumber\\
&&+\frac{\Omega'}{2}(\ket{11}\bra{\psi_+}+{\rm H.c.}).
\end{eqnarray}
%H_{eff}=\left(%
%\begin{array}{cc}
 % 0 & \Omega'/2 \\
  %\Omega'/2 & -\Omega'^{2}/4(V_{\rm {XX}}-2V_{\rm{F}}) \\
 %\end{array}%
%\right).
%\end{equation}
However, within the condition of Eq.~\ref{condition1} the magnitude of the first term is very small compared to that of the other terms. Therefore, the eigenstates of $H_{\rm eff}$ are given approximately by
$2^{-1/2}\{|11\rangle+|\psi_{+}\rangle\}$ and $2^{-1/2}\{|11\rangle-|\psi_{+}\rangle\}$, with respective eigenvalues of
$\Omega'/2$ and $-\Omega'/2$.

Transforming back to the lab frame leads to the following evolution of the initial state $\ket{11}$:
{\setlength\arraycolsep{0.0em}
\begin{eqnarray}\label{timeevolvelab}
|11\rangle&{}\rightarrow{}&\cos\left[\frac{1}{2}\int_0^T\Omega'(t)\rm{d}t\right]|11\rangle\nonumber\\
&&{+}ie^{-i(\omega_{a}+V_{{F}})}\sin\left[\frac{1}{2}\int_0^T\Omega'(t)\rm{d}t\right]
|\psi_{+}\rangle.
\end{eqnarray}}Hence, for $\int_0^T\Omega'(t)\rm{d}t=2\pi$ we generate the required phase change
$|11\rangle\rightarrow-|11\rangle$.

If the CPHASE gate is to work, we must ensure that no basis state, other than
$|11\rangle$, experiences a phase change within the gate time $T$. Returning to the Hamiltonian of Eq.~\ref{spinexH} we
see that the state $|00\rangle$ is completely uncoupled from
any other state, and so it will undergo no phase change during the gate operation.
The states $|10\rangle$ and $|01\rangle$ are not uncoupled and the transitions
$|01\rangle\leftrightarrow|0{X}\rangle$ and $|10\rangle\leftrightarrow|{X}0\rangle$ are possible.
Taking the $\{|01\rangle,|0{X}\rangle\}$ subspace as an example (the $\{|10\rangle,|{X}0\rangle\}$ subspace
is entirely equivalent for two identical dots) and moving into the rotating frame as above, we may write
\begin{equation}\label{Hsub010Xrot}
H'=-V_{F}\ket{0X}\bra{0X}+\frac{\Omega}{2}(\ket{01}\bra{0X}+{\rm H.c.})
\end{equation}
%H'=\left(%
%\begin{array}{cc}
 % 0 & \Omega/2 \\
  %\Omega/2 & \omega_{a}-\omega_{l} \\
% \end{array}%
%\right)=\left(%
%\begin{array}{cc}
 % 0 & \Omega/2 \\
 % \Omega/2 & -V_{\rm{F}} \\
% \end{array}%
%\right),
%\end{equation}
when we insert the laser frequency $\omega_{l}=(\omega_{a}+V_{{F}})$.
Therefore, under the condition
\begin{equation}\label{condition2}
|\Omega|/2 \ll |V_{{F}}|
\end{equation}
the initial state $|01\rangle$ will experience no phase
shift due to the laser~\cite{notes}.

\begin{figure}[t]
\centering
\includegraphics[width=3.0in,height=3.0in]{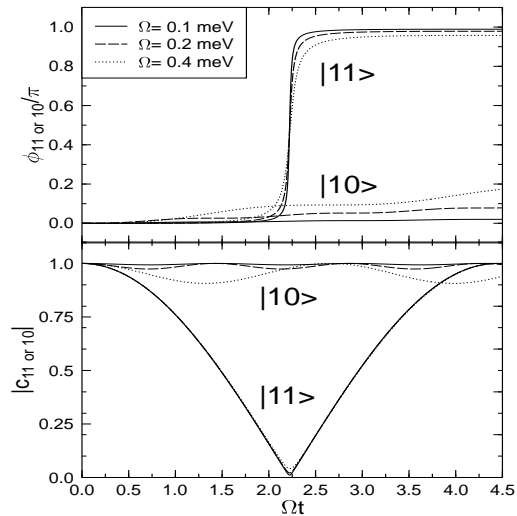}
\caption{Phase (top) and amplitude (bottom) of states $|10\rangle$ and $|11\rangle$ during the CPHASE gate operation,
with $V_{{F}}=0.85$~meV, $V_{{XX}}=5$~meV, and $\omega_a=2$~eV.}
\label{twoqubit}
\end{figure}

In order to estimate the characteristic timescale of our CPHASE gate,
we consider typical interaction strengths $V_{{F}}=0.85$~meV~\cite{nazir04} and
$V_{{XX}}=5$~meV~\cite{biolatti02,lovett03}. Conditions Eq.~\ref{condition1} and Eq.~\ref{condition2}
imply that $\Omega\sim0.1-0.2$ meV ($\Omega'\sim0.14-0.28$ meV) at its maximum for a high fidelity operation
(in the region of $95\%-99\%$). This corresponds to a gate implementation time $T\sim15-30$ ps for a square pulse,
sufficient for many such operations to be performed within measured low-temperature exciton dephasing
times~\cite{borri01,birkedal01,bayer02}.
%Although such fidelities will not be sufficient to perform scalable,
%fault-tolerant quantum computation, they are on a par with the state-of-the-art for other experiments.
In Fig.~\ref{twoqubit} we show a numerical simulation of Eq.~\ref{spinexH} over a complete gate cycle, demonstrating
the suppression of unwanted phase on the state $|10\rangle$ as the ratio $\Omega/V_{{F}}$ decreases.

The CPHASE operation outlined above, while being conceptually simple and the most straightforward gate to implement
experimentally, places stringent conditions on the system dynamics.
However, it is possible to relax the condition of Eq.~\ref{condition2} through the use of single-qubit operations and by
redefining the phase accumulated on state $|11\rangle$ as $|11\rangle\rightarrow e^{i\theta}|11\rangle$~\cite{calarco03}.
Here, ${\theta=\phi_{00}-\phi_{01}-\phi_{10}+\phi_{11}}$ where $\phi_n$ is the phase change of $|n\rangle$ during the gate
operation. Further, we can design the experiment so that no population leaks out of the computational basis after the
completion of the CPHASE gate (for example, by making sure that the periods of evolution of the other qubit states are
commensurate with the $\ket{11}$ evolution). In this case, we find that
for $\theta=\pi$ we recover the simple CPHASE gate. Thus Eq.~\ref{condition2} is now no longer
necessary, and we need only satisfy the condition of Eq.~\ref{condition1}, which may be less restrictive~\cite{lovett03}.

To complete our proposal for a universal set of gates, we now move to single-qubit operations.
Optically, the most straightforward of these is a $Z$ rotation, where a relative phase is accumulated
between $\ket{0}$ and
$\ket{1}$.
%Considering an arbitrary initial state on a single dot, and
If we exploit Pauli blocking as
above, a $\pi$-pulse of $\sigma^{+}$ light tuned to the exciton resonance gives us the transformation
\begin{equation}\label{zrot}
a|0\rangle+b|1\rangle\rightarrow a|0\rangle+b|{X}\rangle.
\end{equation}
If we now allow the system to evolve freely for a time $T_{s}$, and then deexcite via a further $\pi$-pulse, we obtain
the state $a|0\rangle+b\exp(-i\omega_aT_s)|1\rangle$. Here, $\omega_a$ is the energy difference between
$|{X}\rangle$ and
$|1\rangle$, and the final state corresponds to a rotation of the input state by $\phi=\omega_aT_s$ about the $z$ axis
of the Bloch sphere.

Natural size and composition fluctuations in self-assembled dot samples allow for energy selective addressing of individual
QDs, and we must be able to move two coupled dots in and out of resonance in order to perform both single- and two-qubit
manipulations. We propose the use of an inhomogeneous external electric field to enable us to move between the nonresonant
and resonant cases on a timescale which is short compared to typical spin decoherence times~\cite{hanson03,golovach03}.
Field gradients of approximately $20$ (MV/m)/$\mu$m have been obtained experimentally~\cite{rinaldi01} with Stark shifts of
about $2$ meV seen in a field of $0.2$ MV/m in the same experiment. Therefore, for adjacent dots spaced by $5$ nm we can
reasonable expect an energy selectivity of $1$ meV. As any laser resonant with the $|1\rangle\rightarrow|{X}\rangle$
transition on one dot will behave as a detuned laser on a second nearby dot, we may estimate the fidelity ${\cal F}$
of avoiding unwanted transitions in the second dot from
%\begin{equation}\label{fidelity}
${\cal F}=1-(\Omega^2/\delta^2),$
%\end{equation}
where $\delta$ is the laser detuning. Hence, for $\delta=1$ meV, to achieve a $\pi$-pulse with $99\%$ fidelity requires
an operation time of $T_s\sim20$ ps (reducing to $T_s\sim7$ ps for $90\%$ fidelity). As the gate time scales as
$1/\delta$, it would be significantly reduced in a well optimized experiment.

\begin{figure}[t]
\centering
\includegraphics[width=3.0in,height=3.5in]{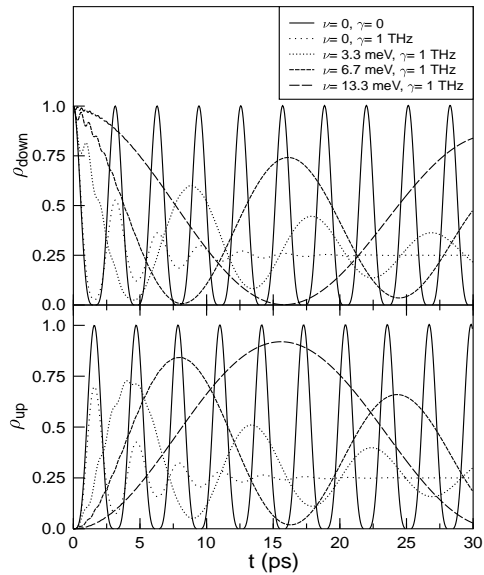}
\vspace{-1cm}
\caption{Population of the state of a single qubit during the proposed $X$ rotation gate induced by a two-laser
Raman transition. Various values of the exciton decay rate $\gamma$ and laser detuning $\nu$ are shown with
$\omega_a=2$~eV,
and $\Omega=1.33$~meV for both lasers.}
\label{onebit}
\end{figure}

Optically induced single spin rotations about a second axis (for example about the $X$ axis) are a more difficult
proposition. Raman transitions involving the light-hole levels $|m_z^{lh}=\pm1/2\rangle$ and a single exciton within a
QD have been proposed as a means to achieve direct optical transitions between the spin states $|0\rangle$ and
$|1\rangle$~\cite{pazy03, chen04}. Unfortunately, as the light-hole states are not hole ground states in GaAs based QDs they
suffer from extremely short decoherence times. Pazy {\it et al.}~\cite{pazy03} propose the use of II-VI semiconductors
to shift
the light-hole energy levels to become ground states. However, since GaAs based QDs are the current state-of-the-art we
would like to be able to implement universal single-qubit rotations in such systems. We therefore adapt the scheme of
Ref.~\cite{pazy03} by using detuned lasers to induce Raman transitions between $|0\rangle$ and $|1\rangle$ whilst exciting
very little population to the fast decaying light-hole states. Figure~\ref{onebit} shows a numerical simulation which
demonstrates that the fidelity of the gate can be improved by detuning the lasers, but at the cost of a longer gate time.
The simulations of Fig.~\ref{onebit} use an exciton decay time $\tau_{X}$ of 1~ps, which is a very pessimistic estimate
but is useful since it clearly demonstrates the detuning effect.
If we take $\tau_{X} = 10$~ps, which is still much shorter
than typical heavy-hole exciton dephasing times~\cite{borri01},
a $\pi$ pulse
fidelity of 99\% can be achieved in 16~ps. It is sufficient for universal quantum computing
to be able to perform this kind of gate globally, so long as the
$Z$ gate, which we have described above, is qubit selective.

An important feature of our scheme is that
the inherent coupling between optical transitions and spin states is just what is required for quantum measurements.
This can proceed through spin-dependent resonance fluorescence, with or without shelving to a metastable
level~\cite{shabaev03}. This
technique is ideally a projective measurement and so could also be used for state preparation.

To summarize, we believe that the scheme outlined above allows for the realization, with available technology, of a
CPHASE gate between two spins in adjacent quantum dots, and for arbitrary single-qubit manipulations.
As our
coupled-spin quantum logic gate requires the excitation of single exciton states only, it has a major advantage in the
need for only one simple laser pulse
to couple the qubits.
Furthermore, we have shown that both types of gate may be performed to a
good fidelity
with the current semiconductor and laser technology; these gates therefore provide
an immediate route to some of the less demanding applications of quantum processors~\cite{dur99}.
Recent work on full-scale fault tolerant
quantum computation (FTQC)~\cite{steane03} also indicates that our favourable ratio of exciton gate decay to spin
qubit decoherence times could greatly decrease the threshold fidelity required for FTQC; estimates
of this threshold are now around 99\%~\cite{reichardt04}.
We therefore expect that it will be possible to use our scheme to build a full-scale quantum processor as
quantum dot growth and characterization techniques progress.

A. N. and B. W. L. are supported by EPSRC (GR/R66029/01).
S. D. B. acknowledges support from the E.U. NANOMAGIQC project
(Contract no. IST-2001-33186).
We thank E.~Pazy, D.~A.~Williams and W.~J.~Munro for useful and stimulating discussions.
\bibliography{spinex}

\begin{thebibliography}{19}
\bibitem[{notes()}]{notes}
\eprint{Though we have considered the specific case of dots with equal exciton energies, a CPHASE gate
can still be performed if the dots are nonresonant, provided that the laser tuning and intensity are
suitably modified. In this case, the gate may take longer to perform.}
\end{thebibliography}
% Produces the bibliography via BibTeX.

\end{document}